\documentclass[aps,pra,twocolumn,superscriptaddress]{revtex4}

\usepackage{amsmath,amssymb}
\usepackage{xcolor,soul}
\usepackage{hyperref,graphicx}

\newcommand{\ket}[1]{|#1\rangle}

\newcommand{\figref}[1]{Fig.~\ref{#1}}

\newcommand{\fisicarm}{Dipartimento di Fisica, Sapienza Universit\`{a} di Roma, Piazzale Aldo Moro 5, I-00185 Roma, Italy}
\newcommand{\fisicami}{Dipartimento di Fisica, Politecnico di Milano, Piazza Leonardo da Vinci 32, I-20133 Milano, Italy}
\newcommand{\ino}{Istituto Nazionale di Ottica, Consiglio Nazionale delle Ricerche (INO-CNR), Largo Enrico Fermi 6, I-50125 Firenze, Italy}
\newcommand{\ifn}{Istituto di Fotonica e Nanotecnologie, Consiglio Nazionale delle Ricerche (IFN-CNR), Piazza Leonardo da Vinci 32, I-20133 Milano, Italy}

\begin{document}

\title{Integrated optical waveplates for arbitrary operations\\on polarization-encoded single-qubits}

\author{Giacomo Corrielli}
\affiliation{\ifn}
\affiliation{\fisicami}

\author{Andrea Crespi}
\affiliation{\ifn}
\affiliation{\fisicami}

\author{ Roberto Osellame}
\email{roberto.osellame@polimi.it}
\affiliation{\ifn}
\affiliation{\fisicami}

\author{Riccardo Geremia}
\affiliation{\fisicami}

\author{ Roberta Ramponi}
\affiliation{\ifn}
\affiliation{\fisicami}

\author{Linda Sansoni}
\affiliation{\fisicarm}

\author{Andrea Santinelli}
\affiliation{\fisicarm}

\author{Paolo Mataloni}
\affiliation{\fisicarm}
\affiliation{\ino}

\author{Fabio Sciarrino}
\email{fabio.sciarrino@uniroma1.it}
\affiliation{\fisicarm}
\affiliation{\ino}

\begin{abstract}
Integrated photonic technologies applied to quantum optics have recently enabled
a wealth of breakthrough experiments in several quantum information areas. Path encoding was initially used to demonstrate operations on single or multiple qubits. However, a polarization encoding approach is often simpler and more effective. Two-qubits integrated logic gates as well as complex interferometric structures have been successfully demonstrated exploiting polarization encoding in femtosecond-laser-written photonic circuits. Still, integrated devices performing single-qubit rotations are missing. Here we demonstrate waveguide-based waveplates, fabricated by femtosecond laser pulses, capable to effectively produce arbitrary single-qubit operations in the polarization encoding. By exploiting these novel components we fabricate and test a compact device for the quantum state tomography of two polarization-entangled photons. The integrated optical waveplates complete the toolbox required for a full manipulation of polarization-encoded qubits on-chip, disclosing new scenarios for integrated quantum computation, sensing and simulation, and possibly finding application also in standard photonic devices.
\end{abstract}

\maketitle

Photons are ideal carriers of quantum information: faint interactions with the environment make them almost immune to decoherence. In particular, the two levels associated to the polarization degree of freedom are advantageously exploited to encode qubits. In fact, entangled-photon sources are available, and many quantum information protocols and experiments have been developed for this encoding \cite{mattle1996dce,lanyon2010tqc,ma2012qt143,sansoni2012tpb, ma2012edc, crespi2013ale, gundogan2012qsp}.
To implement arbitrary operations on qubits, or equivalently to achieve universal quantum computation, two kinds of elementary linear logic gates are sufficient \cite{knill2001sfe}: a two-qubit gate, such as the CNOT gate, and single-qubit rotators. For polarization-encoded photons, two-qubit gates require different spatial modes to interact and are experimentally realized by polarization-dependent beam-splitters. On the other hand, single-qubit operations require an interaction between the two polarization modes belonging to the same spatial mode and are experimentally implemented by optical waveplates.

Proof-of-principle demonstrations of photonic qubits interactions and transformations have been realized by bulk-optics equipment \cite{obrien2009pqt, okamoto2010rkc}. However, building more complex networks of logic gates encounters severe size and stability limitations: an integrated-optics approach allows to overcome these limitations and has enabled a growing number of experiments in the recent years \cite{politi2009iqp,politi2009sqf,obrien2009pqt,tanzilli2012oge,crespi2013ale,spring2013bsp,broome2013pbs,tillmann2013ebs,crespi2013imi}. In particular, a waveguide-based architecture able to support polarization-encoded qubits has been recently developed, exploiting the femtosecond laser waveguide writing technology \cite{sansoni2010pes}.
Femtosecond laser writing \cite{dellavalle2009mpd} exploits non-linear absorption of focused ultrashort pulses to create localized regions with increased refractive index in the volume of dielectric substrates. Translation of the sample at constant speed under the laser beam allows to literally draw three-dimensional waveguiding paths with the desired design. The low birefringence of the fabricated waveguides, when certain irradiation parameters are used, makes them ideal for the propagation of polarization encoded qubits with no loss of coherence \cite{sansoni2010pes}. Polarization dependent or independent directional couplers have been recently demonstrated by this technology, bringing to the realization of a two-qubit CNOT gate \cite{crespi2011ipq}, as well as complex interferometric structures for quantum simulation \cite{sansoni2012tpb, crespi2013ale}.

As a matter of fact this technological platform is not complete, since single-qubit rotations have not been demonstrated yet in integrated optical devices for polarization encoding. Indeed, integrated waveplates with arbitrary axis direction, which effectively allow one to rotate arbitrarily the polarization state of a propagating photon, have never been reported by femtosecond laser writing. Fernandes et al. showed in recent works \cite{fernandes2011flw,fernandes2012sib} that the birefringence of femtosecond-laser-written waveguides can be finely tuned in fused silica substrate and, in particular, they demonstrated highly controlled integrated linear retarders. However, in those works the optical axis of the birefringent waveguides was fixed along the vertical or horizontal direction. Incidentally, one can note that producing integrated waveplates with arbitrary optical axis direction is a challenging task also with other conventional waveguide fabrication technologies, because these components require altering the waveguide symmetry in 3D, which is not easy with planar lithographic techniques \cite{dai2012ptf}.

\begin{figure*}
\includegraphics[width=17cm]{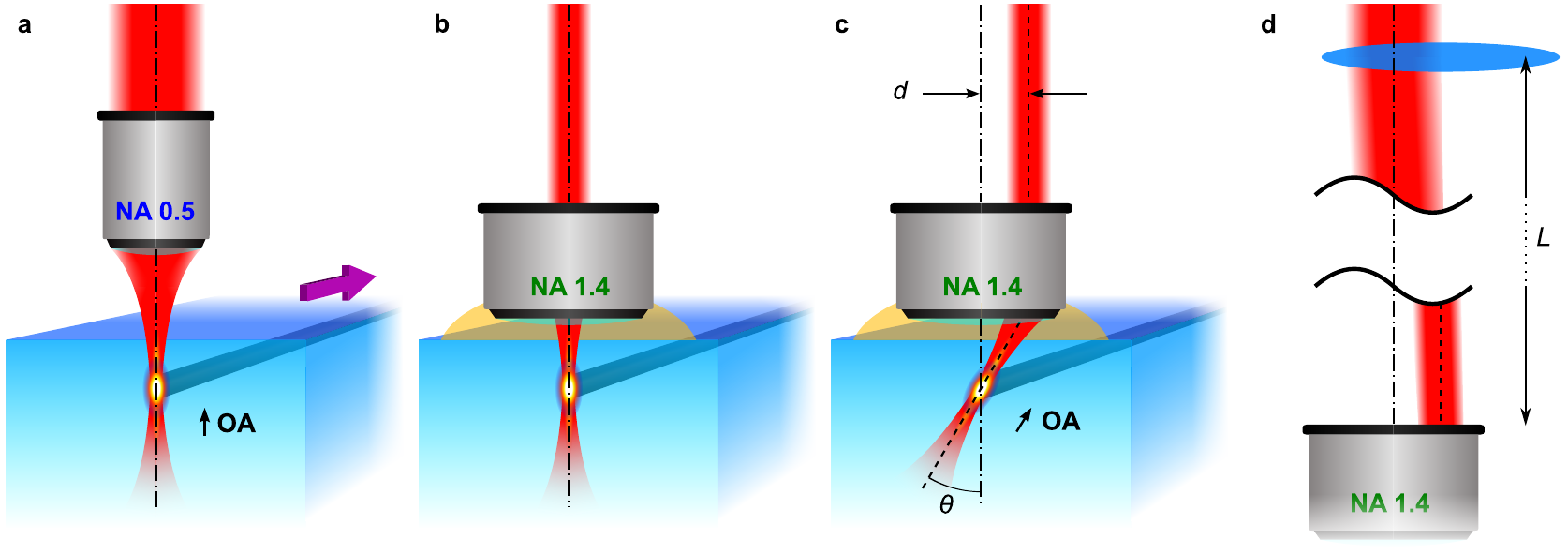}
\caption{\textbf{Femtosecond laser writing of tilted integrated waveplates.} Conceptual scheme of the new method enabling the direct writing of optical waveguides acting as integrated waveplates with tilted axis. \textbf{a}, Traditional writing scheme with a moderate $NA$ focusing objective; the symmetry of the writing layout creates birefringent waveguides with the optical axis (OA) aligned as the writing beam direction; actual waveguide writing is performed by a transversal translation of the glass sample (indicated by the purple arrow). \textbf{b}, Equivalent waveguides can be created by underfilling a high-NA oil-immersion objective. \textbf{c}, Offsetting the writing beam before the objective results in waveguide writing with an inclined laser beam; the resulting waveguide has an optical axis tilted by an angle $\theta$ that depends on the amount of offset $d$ of the writing beam with respect to the objective axis. \textbf{d}, Reduced beam size and offset at the objective aperture is achieved by a small transversal shift of a long-focal lens placed at a distance $L$ from the focusing objective (see Methods for details).}
\label{fig:technique}
\end{figure*}

In this work, we develop a novel method to fabricate integrated waveguide waveplates with tilted axis, by femtosecond laser writing. As we will show, our method requires an extremely simple experimental apparatus. We prove the potential of this technology to implement arbitrary polarization-qubit rotations by realizing and successfully testing a compact device for quantum state tomography {\cite{jame01pra}} of two polarization-entangled photons. 

Femtosecond-laser-written waveguides are slightly birefringent (typical values range in $b=10^{-5}\div 10^{-4}$), i.e. they support guided modes with different propagation constants for two orthogonal polarization states. The microscopic causes of this birefringence may be diverse depending on the substrate material and the irradiation conditions: ellipticity of the waveguide cross-section \cite{snyder1986ofa}, mechanical stresses in the modified region\cite{bhardwaj2004sfl}, or laser-induced intrinsic birefringence aligned according to the writing beam polarization \cite{bricchi2004fbn}. We experimentally verified that in the boro-aluminosilicate glass, which we use to produce integrated quantum photonic circuts \cite{sansoni2010pes}, the polarization direction of the writing laser beam does not affect the waveguide birefringence. Thus, in this glass, we expect only the first two mechanisms to be effective. In this view the waveguide writing process presents an evident symmetry constraint: considering a Gaussian writing beam, a static modification will have a rotational symmetry with respect to the writing beam direction. Hence, independently of the specific microscopic causes of the birefringence, the optical axis will be parallel to this direction.

In the vast majority of the waveguide writing experiments reported in the literature, the laser beam direction is orthogonal to the substrate surface and the writing is performed by scanning the beam transversely to its direction (\figref{fig:technique}a). The waveguide optical axis is therefore typically orthogonal to the sample surface. According to this discussion, the easiest way to tilt the waveguide optical axis would be to rotate the writing beam of an angle $\theta$ with a rotation axis parallel to the waveguide direction. In this way the waveguide cross-section, together with its optical axis, would rotate of the same angle $\theta$. However, a macroscopic rotation of the glass with respect to the writing laser beam, or viceversa, would not be a convenient solution. In fact, in order to avoid large translations of the writing beam inside the glass sample at every rotation $\theta$, a hard-to-obtain resolution in setting the position of the rotation axis would be required. In addition, light refraction at the tilted glass interface would further complicate the fine positioning of the writing focal spot. Finally, the angle-dependent aberrations and Fresnel reflections of the writing beam across the tilted glass interface would make it necessary to re-optimize the laser irradiation parameters for each tilting angle.

We therefore developed a technique to obtain a birefringent waveguide with tilted optical axis without the drawbacks of a macroscopic rotation of the sample with respect to the writing laser beam. Typical waveguide writing is performed with moderate numerical aperture (NA) microscope objectives (e.g. NA~=~0.5, \figref{fig:technique}a) to achieve the required waveguide size in a single scan. Our new method to tilt the waveguide optical axis is based on a high-NA oil-immersion objective (typical NA=1.4, \figref{fig:technique}b) and a writing laser beam with a diameter smaller than the objective aperture. The reduced dimension of the laser beam with respect to the objective aperture is responsible for a reduced effective NA of oil-immersion objective. An optimization of the beam size can yield an effective NA producing the same waveguides one would obtain with the dry, moderate-NA objective (\figref{fig:technique}a,b).

If the laser beam impinges upon the center of the objective, the beam propagates (orthogonally to the substrate surface) without changing its direction. If the laser beam impinges on the objective aperture in an off-center way, the focus position is not altered, but the beam propagates in the substrate at an angle $\theta$ that depends on the offset distance on the objective (\figref{fig:technique}c). The fabricated waveguide will then have a tilted cross-section, resulting in a rotated optical axis. One can observe that, being the focus position independent from the beam offset at the aperture (which is true for any commercial, aberration-compensated, oil-immersion microscope objective), the position of the fabricated waveguide is in principle always the same, for any angle $\theta$. In addition, since the objective remains always orthogonal to the substrate surface, the aberrations are well compensated, while the presence of the immersion oil quenches the Fresnel reflections at the glass interface. This means that the pulse energy delivered at the focal spot is always the same for any $\theta$ and no adjustment is required.

Tilting of the writing beam inside the glass requires transversely offsetting the beam before the focusing objective. This could be effectively done with a pair of motorized mirrors. In our setup, however, we implemented a very simple variant, based on a single, long-focal  lens mounted on a linear translator (\figref{fig:technique}d). The lens, mounted before the focusing objective, has the twofold effect of reducing the beam diameter to achieve the reduced effective NA and, at the same time, to shift the incidence point on the objective aperture by a small transverse translation of the lens. Actually, this is not a pure translation of the beam impinging on the objective, since it also adds an angular deviation. The latter deviation is however very small given the long focal length and the limited objective aperture. A careful analysis showed that the angular deflection introduced by the lens results in a shift in the writing focal point in the order of a few $\mu$m. This shift is very reproducible and can be easily compensated during the waveguide writing process. Details on the experimental setup can be found in the Methods section and in the Supplementary Information.

\begin{figure}
\includegraphics[width=8.5cm]{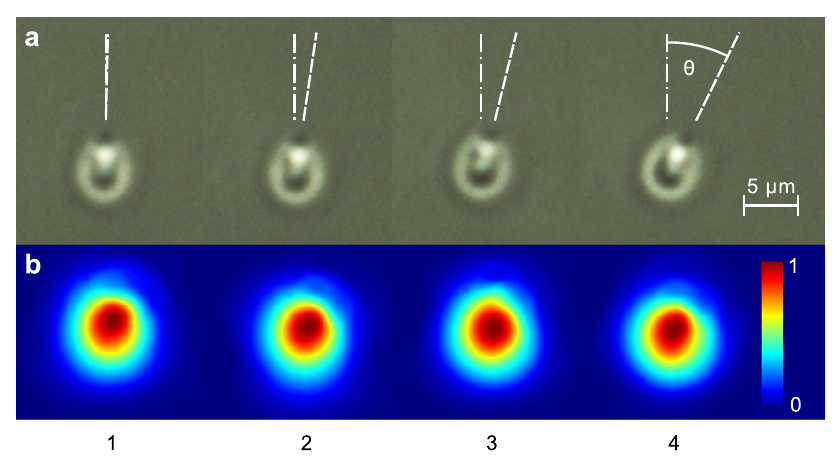}
\caption{
\textbf{Experimental characterization of tilted waveguides.} \textbf{a}, End-view microscope images of waveguides (labeled 1-4) showing a progressive rotation of the cross-section as the offset $d$ of the writing beam is increased. \textbf{b}, Near-field profiles of the waveguide modes at different rotation angles. Quantitative measures are reported in Table I. 
}
\label{fig:tiltedWG}
\end{figure}

\begin{table*} 
	\sf{	
	\centering	
	\begin{tabular}{|c||c|c|c|c|}
		\hline		
		\textit{WG n$^\circ$} & 1 & 2 & 3 & 4\\
		\hline\hline	
		\textit{d} & 0.0~mm & 0.26~mm & 0.75~mm & 0.92~mm\\
		\hline
		$\mathsf{\theta_{wg}}$ & 0$^\circ$ & 10$^\circ$ & 16$^\circ$ & 27$^\circ$ \\
		\hline
		$\mathsf{\theta_{OA}}$ & 1$^\circ$ & 9$^\circ$ & 15$^\circ$ & 25$^\circ$ \\
		\hline
		\textit{b} & 2.03$\times$10$^{-5}$ & 1.97$\times$10$^{-5}$ & 2.04$\times$10$^{-5}$ & 1.99$\times$10$^{-5}$ \\
		\hline
		\textit{Mode size} & 7.3$\times$7.7$~\mu$m$^2$	& 7.4$\times$7.6$~\mu$m$^2$ & 7.2$\times$7.5$~\mu$m$^2$ & 7.0$\times$7.4$~\mu$m$^2$ \\
		\hline
	\end{tabular}}
	\caption{\textbf{Relevant parameters of the waveguides with a tilted optical axis.} Data are referred to the waveguides reported in \figref{fig:tiltedWG}: displacement $d$ of the writing beam before the objective (\figref{fig:technique}c), rotation of the waveguide cross-section $\theta_{wg}$ measured from the optical microscope images (\figref{fig:tiltedWG}a), rotation of the waveguide optical axis $\theta_{OA}$ and birefringence value $b$ determined by an optical characterization of the waveguides, dimensions at $1/e^2$ of the guided-mode intensity profile at 800~nm wavelength (\figref{fig:tiltedWG}b).}
	\label{table:tiltedWG}
\end{table*}

As a first experiment we fabricated several waveguides with different tilt angles, produced by different shifts of the long-focal lens before the focusing objective. For each fabricated waveguide we measured the tilt angle of its cross-section, by acquiring images of the waveguide end-facet with an optical microscope (see \figref{fig:tiltedWG}a). To fully characterize the waveguide birefringence we measured the Stokes vector of the output polarization state when launching several different input polarization states; we then retrieved the value of birefringence and the optical axis tilt angle that best fitted the experimental data for each waveguide. As shown in Table I, we found a very good agreement between the tilt angle of the cross section, retrieved by the microscope images and the tilt angle of the optical axis determined by the polarization analysis. This confirms our idea that rotating the waveguide cross-section implies a rotation of the optical axis of the same angle. It should be observed that the maximum tilting angle achievable with this technique is limited by the numerical aperture of the microscope objective. In our set-up, with NA=1.4, we achieved rotations of the optical axis up to 32$^\circ$ with respect to the vertical axis. In addition, we observe that the birefringence value $b$ remains essentially the same at all the different angles with an average $\overline{b}=(2.01 \pm 0.03) \times 10^{-5}$. Finally, we measured the near-field profile of the guided mode at 800~nm wavelength for each waveguide (\figref{fig:tiltedWG}b), noting no relevant differences in the mode size and shape. This confirms that our technique is able to rotate the optical axis with negligible effect on all the other waveguide features. This very important property implies low insertion losses in practical devices, where waveguides with different optical axis direction will be written consecutively. 

\begin{figure}
\includegraphics[width=8.5cm]{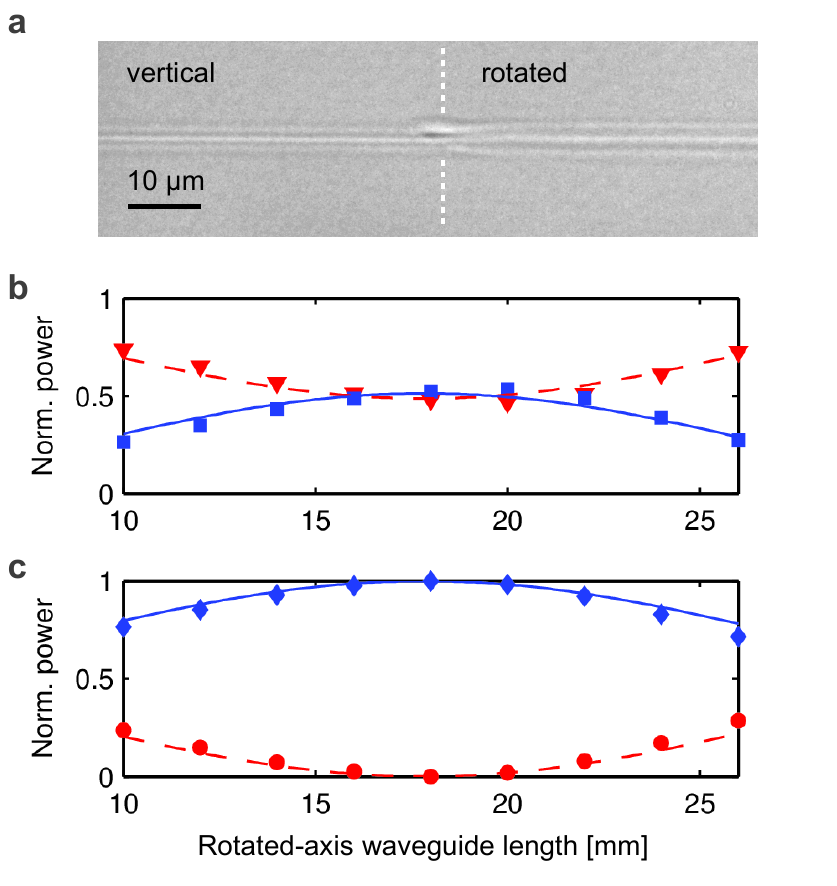}
\caption{
\textbf{Characterization of integrated optical waveplates.} \textbf{a}, Top-view microscope image of a junction between two waveguide sections having different tilt angles. Additional losses introduced by the junction have been measured as 0.3 dB. Several waveplates with a designed optical axis tilt of $\theta=22.5^\circ$ with different lengths have been characterized; for horizontally polarized input light the measured normalized power transferred into (\textbf{b}) the horizontal (red triangles)/vertical (blue squares) polarization states and (\textbf{c}) diagonal (blue diamonds)/antidiagonal (red circles) polarization states is reported. Error bars are omitted since they are smaller than the symbols. Theoretical curves best-fitting the experimental points are plotted and correspond to a $\theta = 21.5 ^\circ$, consistent with the accuracy of our fabrication method in setting the tilt angle, estimated as $\sigma_{\theta} = 1.5 ^\circ$.
}
\label{fig:WP}
\end{figure}
To cascade waveguide segments with different orientations of the optical axis we write the different segments in successive scans. This is made possible by a mechanical shutter synchronized with the motion of the translation stages; the shutter enables the irradiation only in the regions where waveguide segments with a given optical axis tilt have to be fabricated. Then the long-focal lens is translated to a new position and the segments with a different optical axis tilt are produced.  Figure~\ref{fig:WP} shows a microscope top-view image of the junction between two waveguide segments with different optical axis direction; a perfect alignment as well as a hardly noticeable discontinuity can be appreciated. To test our capability to fabricate effective integrated waveplates we realized waveguides comprising two segments, the first with vertical optical axis, the second with optical axis tilted by 22.5$^\circ$. Keeping constant the overall length of the waveguide, we varied the length of the two segments. We characterized the devices by injecting horizontally and vertically polarized laser light into the segment with a vertical optical axis and by measuring the projections on the vertical (V), horizontal (H), diagonal (D) and antidiagonal (A) polarization states of the output (see, for example, \figref{fig:WP}b and c for H polarized input state projected onto the H/V and D/A bases, respectively). The first waveguide segment has actually no influence on the vertical or horizontal input polarization states, while the second one introduces a polarization rotation. The induced polarization rotations fit very well with the theoretical behaviour calculated for the different lengths of the segment with rotated axis. At a suitable length of about 18 mm of the second segment a clear half-waveplate operation is observed: more than 99\% of the incoming H polarized light is converted into D polarization state. Coherently, no light is found in the A state, while an even splitting of light is found in the H/V basis. 

\begin{figure*}
\includegraphics[width=18cm]{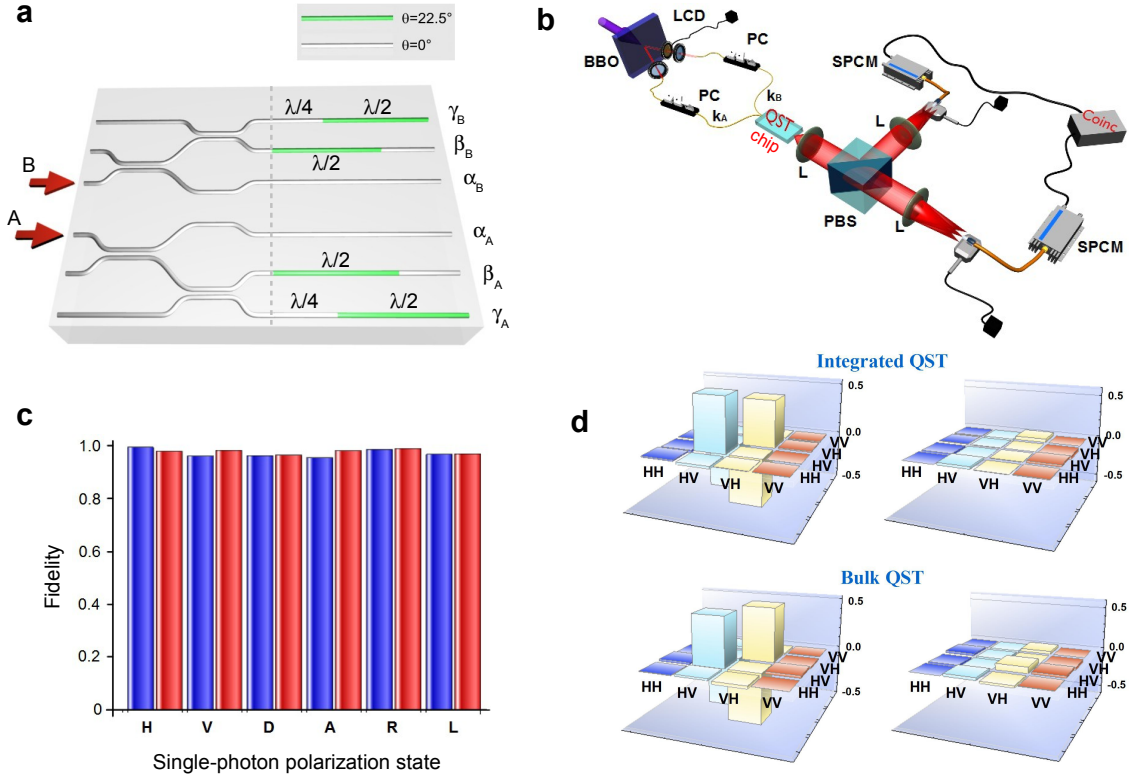}
\caption{
\textbf{Quantum state tomography (QST) in an integrated chip.} \textbf{a}, Scheme of the integrated device (QST chip) for quantum tomography of a two-photon state: it is composed of two specular circuits, one for photon A and one for photon B; these circuits, by means of integrated waveplates of suitable length and tilt, prepare the two-photon state in order to have it projected on the horizontal (H)/vertical (V), diagonal (D)/antidiagonal (A) and circular left(L)/circular right(R) bases by a single polarizing beam splitter (PBS) placed afterwords (see setup in panel \textbf{b}). {\textbf{b}, Scheme of the experimental setup for entantangled states tomography (for details see Methods).} \textbf{c}, Fidelities for quantum tomography of different single-photon polarization states; blue and red bars represent the results obtained by inputting the same polarization state in A and B, respectively. {Error bars are omitted because too small compared to the axis scale.} \textbf{d}, {Real (left) and Imaginary (right) part of the} reconstructed density matrix for a polarization-entangled Bell state ($\Psi^-$); comparison is shown between the measurement made with the integrated chip and that performed by a standard QST setup with all bulk optical elements.}
\label{fig:QST}
\end{figure*}
As a probing benchmark of our technology, we fabricated the complex device represented in \figref{fig:QST}a, which is an integrated optical analyser of two-photon polarization states. The device is composed of two twin parts, one per each photon (A and B). Each part comprises a network of polarization independent couplers (we employed the three-dimensional architecture devised in Ref.~\cite{sansoni2012tpb}) that equally split the input light into three arms. The $\alpha$ arm consists in a waveguide with vertical birefringence axis. The $\beta$ arm terminates with a waveguide segment operating as a half-waveplate, tilted by 22.5$^\circ$ (the remaining segment with tilt 0$^\circ$ does not have any effect on the polarization measurement since its optical axis is aligned to the polarizing beam splitter placed at the output, see \figref{fig:QST}b). The $\gamma$ arm terminates with two consecutive waveguide segments, acting as a vertically oriented quarter waveplate and a half-waveplate tilted by 22.5$^\circ$. The polarization rotations induced by the input fibers and by the first part of the integrated device (up to the dashed line in \figref{fig:QST}a) are compensated by fiber polarization controllers in each input mode (\figref{fig:QST}b). By cascading to this device a single (bulk) polarization beam splitter it is possible to achieve a parallel analysis of the polarization state of the input light in the three basis (vertical and horizontal, diagonal and antidiagonal, right- and left-handed circular polarization). Hence, a complete tomography of one- and two-photon polarization states can be accomplished. Figure \ref{fig:QST}c,d shows the results of our  measurements with single and two-photon states (technical details in the Methods and in the Supplementary Information). In particular, \figref{fig:QST}c reports the fidelities of the tomographic measurements on single photon states, with an average value of {$0.976\pm0.003$} Figure \ref{fig:QST}d shows the reconstructed density matrix of a polarization-entangled Bell state, which yields a {$0.971\pm0.006$} fidelity to the same measurement performed with a standard bulk-optics quantum state tomography equipment.

In conclusion, we developed a new technique to fabricate waveguide segments acting as integrated waveplates, with precise control of the tilt angle. We demonstrated the capability of our devices to perform arbitrary transformations on the polarization state of light propagating in an integrated waveguide circuit, with both classical and quantum states of light. In particular, an integrated device for on-chip analysis of the polarization state of two entangled photons is provided. The integrated optical waveplate constitutes one of the fundamental components, and the last missing, of an integrated platform aimed at the manipulation of polarization-encoded qubits. 
{The demonstration of this device opens the way to the integration of a broader set of experiments and devices that use photon polarization as the most general degree of freedom for implementing many quantum information processes. In particular, the on-chip control of photon polarization will enable the implementation of on-chip polarization-based quantum metrology \cite{mitchell2004srp} and, if combined with path encoding, the integrated manipulation of hyperentangled photon states \cite{vall09pra}.}

\subsection*{Methods}

\textbf{Femtosecond laser waveguide writing} Our waveguide fabrication setup is based on a regeneratively amplified Yb:based femtosecond laser system (HIGHQLaser FemtoREGEN), with wavelength $\lambda = 1040$~nm, pulse-duration of 400~fs and 960~kHz repetition rate. The beam is focused by a $NA = 1.4$ oil-immersion microscope objective, at 170~$\mu$m depth below the surface of borosilicate glass substrates (Corning EAGLE 2000). Sample translation at constant speed is provided by Aerotech FiberGLIDE high-precision, air-bearing translation stages. To deflect the laser beam and translate its incidence point on the objective aperture, a spheric lens with focal length $f = 50$~cm was used, located at a distance $L = 44$~cm from the objective aperture. This allowed us to reduce the laser spot size to a $1/e^2$ diameter $2 w =$~1.6~mm, which is about 2.8 times smaller than the objective aperture $D =$~4.5~mm. The effective numerical aperture approximately scales by the same factor  with respect to the full value $NA=1.4$, reaching an $NA_{\mathrm{eff}}\sim0.5$, which is comparable to values commonly employed for femtosecond laser waveguide writing. Optimum single-mode waveguides at 800~nm were fabricated with a pulse energy of 210~nJ and translation speed of 20~mm/s. Propagation losses at 800 nm wavelength have been measured as 0.2~dB/cm. A mechanical shutter (Thorlabs SH05) is used to append different waveguide segments written with different optical axis directions.

\textbf{Measurements with quantum light}
Let us refer to Fig. \ref{fig:QST}b. To perform measurements with quantum light, a $\beta$-barium borate (BBO) crystal cut for type II phase matching and pumped by a CW diode laser ($\lambda=$~405~nm), generates polarization-entangled photon pairs in the state $\ket{\psi^-}_{AB}=\frac{1}{\sqrt{2}}(\ket{H}_A\ket{V}_B-\ket{V}_A\ket{H} _B)$ at $\lambda=$~810~nm via spontaneous parametric down conversion \cite{kwia95prl}. Photon pairs are then delivered to the integrated circuit (QST chip) through single-mode fibers. The output modes of the chip are {collected by an aspheric lens (L) and} delivered to a bulk polarizing beam splitter (PBS) which transmits horizontal and reflects vertical polarized light.
Both the transmitted and the reflected modes are collected by multimode fibers mounted on motorized stages.
Photons are then detected by single-photon-counting modules (SPCM), and coincidences between photon pairs are finally recorded.

\textbf{Birefringence compensation in the quantum state tomography measurements}
The waveguides constituting the first part of the fabricated device for quantum state tomography are birefringent, and some birefringence is also present unavoidably in the optical fibers used to inject photons in the chip.
To compensate for the polarization rotations induced by the propagation in these sections of the circuit (fibers and waveguides) we adopted the following strategy. First, we compensated horizontal and vertical polarization by means of standard polarization controllers (PCs) acting on the input fibers, by injecting horizontally polarized light and measuring the state emerging from the $\alpha$ outputs (see \figref{fig:QST}a). Second, we adjusted the phase between horizontal and vertical polarizations with a liquid-crystal controllable waveplate (LC), by injecting into the chip diagonal (circular) polarized light and measuring the state emerging from the $\beta$ ($\gamma$) outputs.

\vspace*{1pt}
\subsection*{Acknowledgements}
This work was supported by PRIN 2009 (Progetti di Ricerca di Interesse Nazionale 2009), the ERC-Starting Grant 3D-QUEST (3D-Quantum Integrated Optical Simulation; grant agreement no. 307783): http://www.3dquest.eu, QWAD (Quantum Waveguides Applications \& Developments), EU-Project CHISTERA-QUASAR

\end{document}


\title{Integrated optical waveplates for arbitrary operations\\on polarization-encoded single-qubits\\- Supplementary information -}

\author{Giacomo Corrielli}
\affiliation{\ifn}
\affiliation{\fisicami}

\author{Andrea Crespi}
\affiliation{\ifn}
\affiliation{\fisicami}

\author{ Roberto Osellame}
\email{roberto.osellame@polimi.it}
\affiliation{\ifn}
\affiliation{\fisicami}

\author{Riccardo Geremia}
\affiliation{\fisicami}

\author{ Roberta Ramponi}
\affiliation{\ifn}
\affiliation{\fisicami}

\author{Linda Sansoni}
\affiliation{\fisicarm}

\author{Andrea Santinelli}
\affiliation{\fisicarm}

\author{Paolo Mataloni}
\affiliation{\fisicarm}
\affiliation{\ino}

\author{Fabio Sciarrino}
\email{fabio.sciarrino@uniroma1.it}
\affiliation{\fisicarm}
\affiliation{\ino}

\maketitle

\section{Description of the polarization state and waveplate operation}

We will recall in this Section some basic formalism describing the polarization state of a photon and the action of a birefringent waveplate, and we will show that by combining the integrated waveplates presented in the Main Text it is possible to implement arbitrary waveplate operations.

The polarization state of a single photon may be described by the state vector:
\begin{equation}
\ket{\Psi} = \alpha \ket{H} + \beta \ket{V}
\end{equation}
where $\ket{H}$ and $\ket{V}$ are two orthogonal polarization eigenstates, corresponding to a linear polarization state parallel to the $x$ axis and $y$ axis respectively. The vector $J = \begin{bmatrix} \alpha\\ \beta \end{bmatrix}$ corresponds to the Jones vector describing the classical polarization state of a coherent beam.

The evolution of the polarization state of a photon propagating through a linear optical element, according to the formalism of the Jones vectors and matrices, is given by a linear matrix operation on the vector $J$:
\begin{equation}
J' = M J
\end{equation}
where $M$ is a $2 \times 2$ matrix describing the optical element. For a generic waveplate (or birefringent medium), having the fast axis at an angle $\theta$ with respect to the $x$ axis and inducing a phase delay $\delta$ between the two polarization components, it holds:
\begin{equation}
M \left(\theta, \delta\right) = \begin{bmatrix}
\cos^2 \theta + e^{i \delta} \sin^2 \theta& \left(1-e^{i \delta} \right) \sin \theta \cos \theta\\
\left(1-e^{i \delta} \right) \sin \theta \cos \theta& e^{i \delta}  \cos^2 \theta + \sin^2 \theta
\end{bmatrix}
\label{eq:WPmatrix}
\end{equation}

Observing the form of the matrix \eqref{eq:WPmatrix} one can easily notice that waveplates with different values of $\theta$ and $\delta$ may be equivalent, in particular:
\begin{align}
M \left(\theta, \delta\right) &= M \left(\theta + \pi, \delta\right) \label{eq:WPeq1}\\
M \left(\theta, \delta\right) &= e^{- i \delta} \cdot M \left(\theta - \dfrac{\pi}{2}, -\delta\right)
\label{eq:WPeq2}
\end{align}
where the common phase factor $e^{- i \delta}$ in the second equivalence  has no influence on the polarization state. This means that a set of waveplates with $\theta \in \left[ -\dfrac{\pi}{4}; \dfrac{\pi}{4} \right)$ and $\delta \in \left[ 0; 2 \pi \right)$ is sufficient to perform every waveplate operation.

One can further show that an even smaller set of waveplates is sufficient, provided that one decomposes the action of a single generic waveplate into the action of several cascaded waveplates of this smaller set. As an example (the decomposition is not unique), one can combine a half-waveplate at an angle $\theta$:
\begin{equation}
M \left( \theta, \pi \right) = \begin{bmatrix}
\cos 2 \theta& \sin 2 \theta \\
\sin 2 \theta& -\cos 2 \theta
\end{bmatrix}
\end{equation} 
a waveplate with phase delay $\delta$ and $\theta=0$:
\begin{equation}
M \left( 0, \delta \right) = \begin{bmatrix}
1 & 0 \\
0& e^{i \delta} 
\end{bmatrix}
\end{equation} 
and again a half-waveplate $M \left( \theta, \pi \right)$ at the same angle $\theta$ of the first one. As can be easily calculated, the combination of these three waveplates gives:
\begin{equation}
M \left( \theta, \pi \right) \cdot M \left( 0, \delta \right) \cdot M \left( \theta, \pi \right) = M \left( 2 \theta, \delta \right)
\label{eq:WPcomb}
\end{equation}
This implies that by combining three waveplates with $\theta \in \left[ -\dfrac{\pi}{8}; \dfrac{\pi}{8} \right)$ and $\delta \in \left[ 0; 2 \pi \right)$ as in \eqref{eq:WPcomb}, and considering the equivalences \eqref{eq:WPeq1},\eqref{eq:WPeq2}, it is possible to reproduce the operation of an arbitrary waveplate belonging to the full set $\theta, \delta \in \left[ 0; 2 \pi \right)$.

It is thus proved that, notwithstanding the limitations in the range of angles achievable by the integrated waveplates reported in the Main Text, it is still possible to reproduce arbitrary waveplate operations.

\section{Details of the waveguide fabrication setup}

\subsection{Positioning of the spheric lens}

As discussed in the Main Text, a spheric lens $SL$ with focal length $F=50$~cm is mounted on a linear translation stage in our waveguide writing setup. This lens is used to obtain both a spot-size reduction of the writing laser beam and an effective transverse displacement of the laser beam itself on the aperture of the high-NA objective.

The effective numerical aperture $NA_{eff}$ at which the objective operates depends on the ratio between the beam diameter $2w$ and the objective entrance diameter $D$. In particular, one can consider that the full numerical aperture ($NA = 1.4$) is achieved if $2w = D$ while for lower values of $2w$ a linear relation holds as a first approximation: $NA_{eff} \sim NA \frac{2w}{D}$. We experimentally characterized the beam spot-size as a function of the distance $L$ from the lens $SL$ and we found that $2w = $1.6~mm for $L = 44$~cm, which gives $NA_{eff} \sim 0.5$ since $D = 4.5$~mm. This value of $L$ was thus fixed as lens-objective distance for all our experiments (see \figref{fig:geometry} of Supplementary Information for illustration of the relevant geometrical parameters).

We can also note that, since the focal length of the spheric lens is much larger than that of the objective ($F>>f$), the resulting focal distance after the objective is practically not affected by the presence of the lens. This means that the beam impinging on the objective can be still considered collimated to a good approximation.

\subsection{Effects of the translation of the spheric lens}

\begin{figure*}
	\includegraphics[width=17cm]{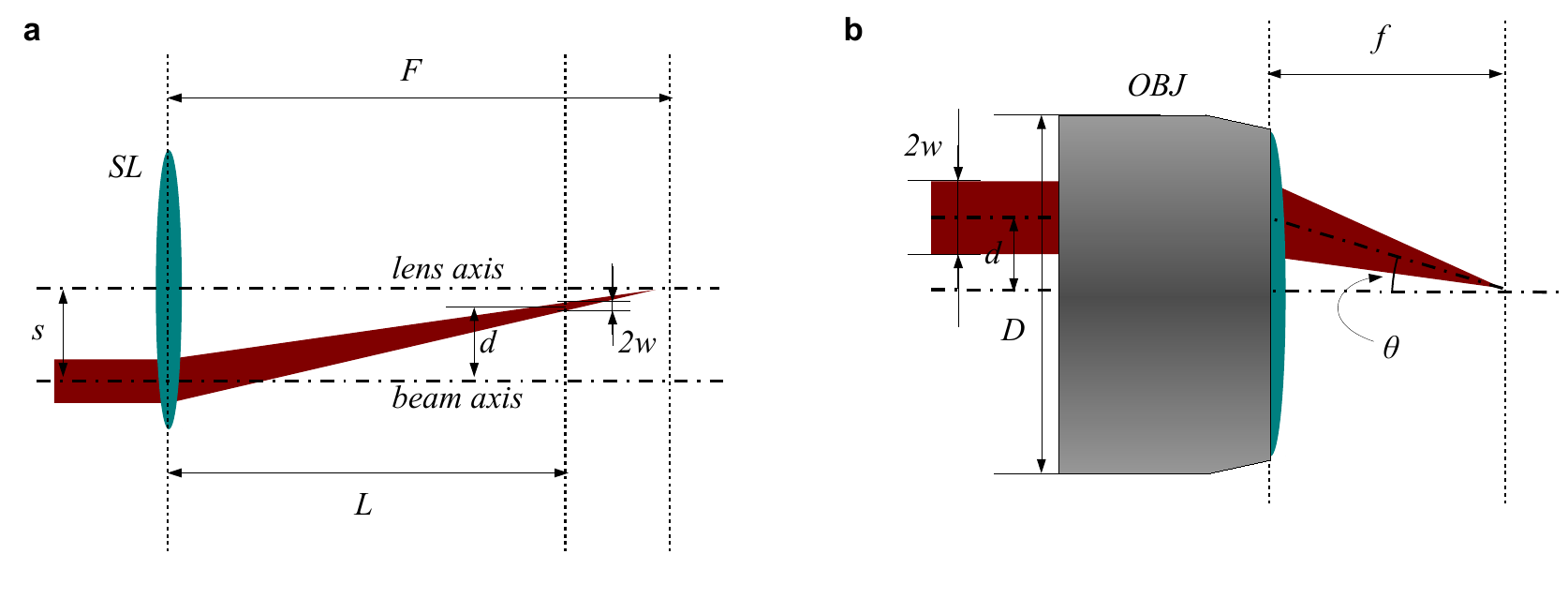}
	\caption{\textbf{a} Geometric scheme representing the beam displacement $d$ induced by a spheric lens $SL$ of focal length $F$, at a distance $L$, for a transverse offset $s$ of the lens itself. \textbf{b} Geometric scheme representing the writing laser beam tilted by an angle $\theta$ when impinging on the objective ($OBJ$) in an off-center position, with a displacement $d$; $f$ is the focal length of the objective and $2w$ is the beam waist diameter.}
	\label{fig:geometry}
\end{figure*}

The displacement $d$ of the beam on the objective aperture, as a function of the transverse displacement $s$ of the lens with respect to the axis of the incoming beam, can be calculated as 
\begin{equation}
d = \frac{L}{F} s
\end{equation}
as it is evident from the geometric construction in \figref{fig:geometry}a.

The angular deflection of the beam impinging on the objective is
\begin{equation}
\theta_D \simeq \frac{d}{L} = \frac{s}{F}
\label{eq:deflection}
\end{equation}
note that in our experiments typically $|s| <$~2~mm thus $\theta_D < 0.2^\circ$. As mentioned in the Main Text, this small deflection $\theta_D$ may cause a small transverse displacement of the focal spot after the writing objective, with respect to the ideal case where the beam impinging on the objective is always parallel to the objective axis. This results in a lateral offset between waveguide segments written with different shifts $s$ (i.e. with different rotations of the optical axis). However, these offsets are highly reproducible and therefore can be effectively compensated by translating the sample of the opposite amount when writing the different segments. In particular, when writing the integrated QST circuit that we present in the main text, an 11.2 $\mu$m lateral offset had to be compensated between the waveguide sections at $0^{\circ}$ and $22.5^{\circ}$.

\begin{figure*}
	\includegraphics[width=7cm]{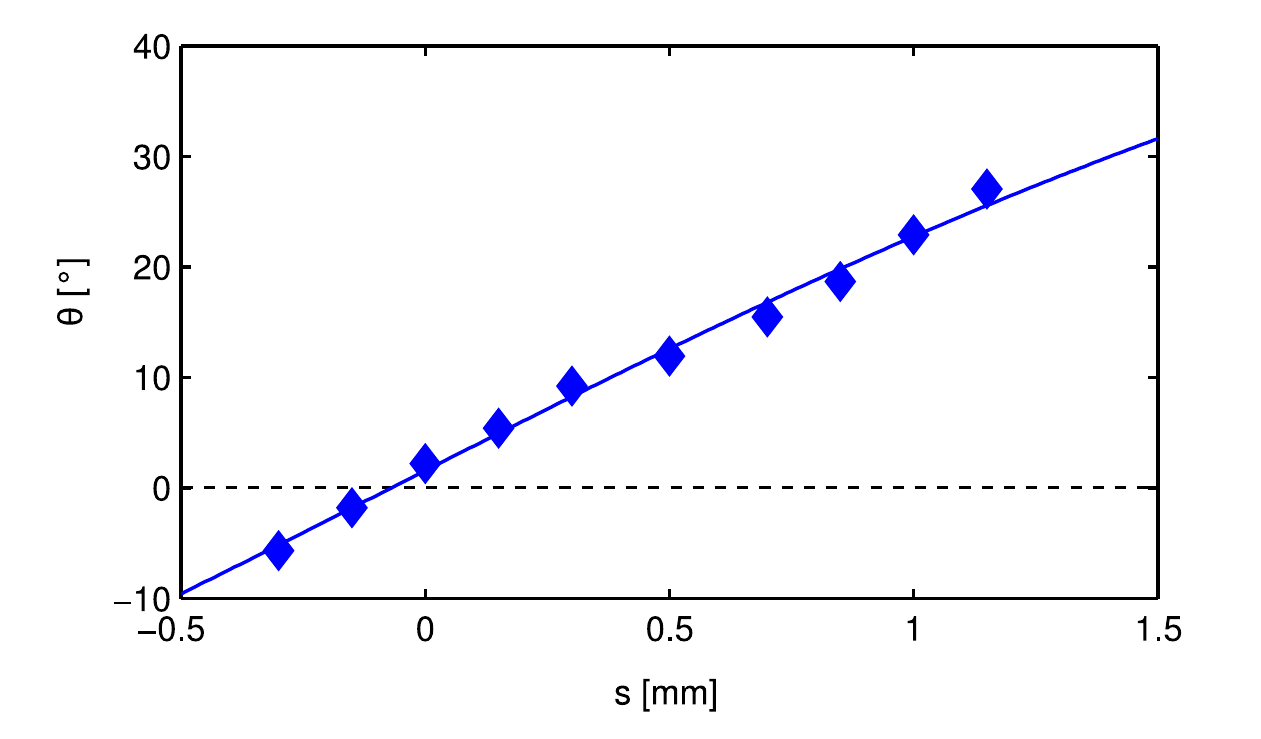}
	\caption{Experimental calibration of the achieved waveguide-axis tilt $\theta$ for a translation $s$ of the spheric lens. Diamonds represent experimental points, while the solid line represents a best-fit with a curve of the form \eqref{eq:fittaAngolo} ($C = 0.39~\mathrm{mm^{-1}}$ and $s_0 = -0.069$~mm) }
	\label{fig:cal}
\end{figure*}

The angle defining the propagation direction of the beam after the objective (having focal length $f$), which determines the tilt angle of the optical axis of the fabricated waveguide, can be also worked out easily by looking at \figref{fig:geometry}b:
\begin{equation}
\theta=\arctan \left( \frac{d}{f} \right) = \arctan \left( s \frac{L}{f F} \right)
\end{equation}
where we have neglected the small angular deviation $\theta_D$, since it produces a second order effect that can be effectively compensated as previously discussed.

In the experimental practice, to accurately calibrate our fabrication process, we fabricated waveguides with different lens displacement $s$ and measured the resulting $\theta$ (see \figref{fig:cal}). The measured points were fitted with a curve of the form:
\begin{equation}
\theta = \arctan \left[ \left(s-s_0 \right) C \right]
\label{eq:fittaAngolo}
\end{equation}
where $C$ and $s_0$ are free parameters. $C$ comprises the different geometrical parameters of the setup, while $s_0$ allows to compensate for non-perfect centering of the spheric lens.

\section{Single-photon quantum state tomography measurements}

In this Section we report on single-qubit quantum-state-tomography measurements realized by means of the integrated-optics device, fabricated by femtosecond laser micromachining.
By starting from pairs generated by spontaneous parametric down conversion in a nonlinear crystal (see Fig. 4b of the Main Text), we generate single photon states: one photon of the pair is used as trigger for coincidence counts, while the other one is encoded in the polarization degree of freedom by means of a polarizing beam splitter, a half- and a quarter-waveplate. 
By this strategy, we prepare single photons in six different states of polarization: horizontal (H), vertical (V), diagonal (D), antidiagonal (A), right- (R) and left- (L) circular and deliver them to the device, which projects each incoming state into the six polarizations H, V, D, A, R, L, as widely described in the Main Text. These projections allowed us to reconstruct the density matrix of the incoming light through quantum state tomography \cite{jame01pra}. These measurements have been performed for photons injected in both modes A and B (see Fig. 4a in Main Text). In Table \ref{tab:fidelities} we report the obtained fidelities, i.e. the overlap, between the expected density matrices and the experimental ones. The average fidelity has been found to be $\mathcal{\bar F}=0.976\pm0.003$. Errors are computed from Poissonian statistics of coincidence counts.

\begin{table*} 
	\sf{	
	\centering	
	\begin{tabular}{|c||c|c|}
		\hline		
		{Input State} & Fidelity A & Fidelity B\\
		\hline\hline	
		H & $0.995\pm0.003$ & $0.980\pm0.003$\\
		\hline
		V & $0.962\pm0.003$ & $0.984\pm0.003$\\
		\hline
		D & $0.964\pm0.006$ & $0.966\pm0.005$\\
		\hline
		A & $0.956\pm0.006$ & $0.982\pm0.004$\\
		\hline
		R & $0.987\pm0.003$ & $0.990\pm0.003$\\
		\hline
		L & $0.969\pm0.004$ & $0.970\pm0.004$\\
		\hline
	\end{tabular}}
	\caption{Fidelities of single qubit quantum state tomography measurements, corresponding to different polarization states: horizontal (H), vertical (V), diagonal (D), antidiagonal (A), right- (R) and left- (L) circular- for both input modes A (central column) and B (right column).}
	\label{tab:fidelities}
\end{table*}